\documentclass[sigconf]{acmart}
\renewcommand\footnotetextcopyrightpermission[1]{} 
\AtBeginDocument{%
  \providecommand\BibTeX{{%
    \normalfont B\kern-0.5em{\scshape i\kern-0.25em b}\kern-0.8em\TeX}}}

\usepackage{listings}

\begin{document}

\title{ASVAAN: Semi-automatic side-channel analysis of Android NDK}

\author{Valerio Brussani}
\email{valerio@nozero.io}

\begin{abstract}
Android is the most popular operating systems for smartphones and is also well-known for its flexibility and security. However, although it is overall considered very secure, there are still some vulnerabilities occasionally discovered that allow getting user sensitive information bypassing security controls and boundaries: among these, side-channel vulnerabilities are a significant concern these days. Although there are several types of side-channel vulnerabilities, ones focused on APIs still represent a great area to explore, which, until now, has often been analysed manually. Only in the latest years, there have been published some automatic solutions which focus on performing automatic scanning of side-channel flaws in Android, created due to the increasing codebase of the operating system; however, they present some limitations. 

This paper introduces a new approach to discover Android NDK side-channel leaks, which at the best of the author knowledge have never been investigated through the usage of automatic or semi-automatic solutions. The approach described in the work, allowed to identify more than 8 new side-channel leaks in several Android NDK functions,which permitted to infer with great accuracy application and websites launches on a victim device. The findings represents the first discovered side-channel leaks in Android NDK functions, and were responsibly disclosed to the Android Security Team of Google.

\end{abstract}



\maketitle
\pagestyle{plain}

\section{Introduction}
Nowadays, Android smartphones are containers of sensitive information, and their security is radically important. Android security relies on two essential concepts: application sandboxing and permission systems, both enforced by the operating system framework.  Although these mechanisms are considered intrinsically secure, there exists a category of vulnerabilities called side-channel attacks, which allow inferring sensitive information related to an event without having appropriate permissions of doing so, by monitoring some influenced physical or logical properties \cite{b1}. 

Side-channels can be of several types, such as micro-architectural, network-based or sensor-based, but one category of specific interest is the one related to Android APIs. Due to a large number of APIs introduced with new Android releases, manually analysing this codebase is a tedious and prone to error task. As a consequence, in recent years, some automatic solutions have been introduced to automatically perform the scanning of Android Framework resources for these types of vulnerabilities \cite{b2}. However, the few available existing options have room for improvements, to increase code coverage and consequently discover new side-channel leaks. Approach to fully automatic or semi-automatic side-channel analysis on mobile devices are very scarce as currently do not exist many projects in literature.{\sc ProcHarvester} only focuses on procfs and only on three events: app starts, website launches and tap actions \cite{b3}. However, since most of the latest years projects are based on leaks of procfs filesystem files, this source has been limited starting from Android 8.0 and for this reason attacks based on it currently results very difficult. On the other hand, {\sc SCAnDroid} which focuses on Android Java APIs covers a limited amount of the available methods, specifically the ones prefixed with certain keywords, and more importantly, Android NDK APIs are completely left aside, which also represent a large codebase to investigate. Furthermore, {\sc SCAnDroid} profiling phase is based only on methods return values and could be extended to include others combined attacks, such as timing attacks \cite{b4} and other custom attacks based on logging of non-original return values. Finally, both the described solutions have been tested on Android 8.0; at the time of writing, the current Android version is 10.0 and at the best of the author knowledge no further automatic side-channel vulnerabilities analysis solution for Android has been implemented since 2018. 
The proposed solution, differently from the previous ones, focuses on Android NDK. It was necessary to implement a different technique to inspect the Android NDK codebase, as it is not possible to use Java reflection. Furthermore, the profiling phase will not only be based on method return values but also on other characteristics customisable by users, as consequence of the findings described in \cite{b4}; therefore, the followed approach allows to implement custom function calls to the Android NDK APIs and other native libraries, by creating a specific function template. This permits to generate custom attacks using other behaviours besides only logging the original return value of a function (e.g. timing attacks or custom return values). 

The aim of the research is to create a new prototype solution, which is complementary to {\sc SCAnDroid}, and increases code coverage by focusing on Android NDK APIs and other native libraries' functions. The research poses the following questions: 

\begin{enumerate}
	\item How do we automatically identify side-channel information leaks in NDK APIs? 
	\item Can we exploit some APIs to get information regarding events of interest? 
	\item Can we combine return values with other functions behaviours or components (such as methods execution time) to identify additional leak sources? 
\end{enumerate}

\begin{table*}[t]
	\begin{center}
		\caption{Side-channel mobile attacks}
		\label{table1}
		\small
		\setlength\tabcolsep{2pt}
		\begin{tabular}{|l|c|c|c|c|}
			\hline
			{\bf Side-channel mobile attacks}   & Public filesystem resources (e.g. procfs) & Sensors data   & Java API return value & Timing Attack \\ \hline
			Website Profiling                   & \cite{b6},\cite{b7},\cite{b4},\cite{b3}                           & \cite{b8}            & \cite{b2}                  & /             \\ \hline
			User info (ex. locations)           & \cite{b4}                                      & \cite{b9},\cite{b10}       & \cite{b2}                   & /             \\ \hline
			Tap \& Input                        & \cite{b3},\cite{b11},\cite{b12}                             & \cite{b13},\cite{b14},\cite{b15} & /                     & /             \\ \hline
			Application launches and foreground & \cite{b4},\cite{b16},\cite{b3},\cite{b12}                         & /              & \cite{b2}                   & /             \\ \hline
			File existence                      & /                                         & /              & /                     & \cite{b4}           \\ \hline
		\end{tabular}
	\end{center}
\end{table*}

\section{Background}
\subsection{Side-channel attacks}
{\em Side-channel vulnerabilities} are currently a hot topic in the computer science research field, also because of the recent discovery of exciting and innovative bugs in popular software and hardware devices. It is considered a very new category of flaw since the first dated side-channel attack was discovered about 20 years ago \cite{b5}. In a nutshell, side-channel attacks exploit unintended leakage of information from hardware or software of a computing device (such as mobile phones, IoT devices and laptops), to infer sensitive information by leveraging an unintended feature of the system.  

One of the most popular types of side-channel attack is applied to cryptographic algorithms, in order to recover information about a used sensitive key, usually by evaluating the amount of time taken to perform a cryptographic operation or by monitoring the electromagnetic emissions \cite{b17}. As described in \cite{b1}, the first types of side-channel attacks required an attacker to be in physical possession of a device in order to leak sensitive information. However, most of the current attack models only have as a prerequisite the installation of a malicious application or the navigation on a mobile website which can execute JavaScript code. In fact, according to that paper, side-channel attacks are divided in local, vicinity and remote ones: local and vicinity attacks require a malicious actor to be near to a victim to exploit a vulnerability, whilst remote attacks can be performed remotely through the installation of an element on the victim device (e.g. an Android application on mobile devices). 

This research is focused on the last type of attack, also known as remote-based side-channel attack. Furthermore, according to \cite{b17}, there are passive and active side-channel attacks: in active attacks, during the exploitation, there is an interaction with the victim component, which affects its state in a certain way, and data is inferred based on its response. On the other hand, in passive attacks, there is no active interaction with the victim device, and sensitive data is inferred passively monitoring the state of the victim device.

\subsection{Side-channel vulnerabilities on mobile devices}
A recent and yet not fully explored type of side-channel attacks are the ones which exploit components of mobile devices. Most of them adhere to the following adversary model \cite{b3}:  
\begin{itemize}
	\item An application which does not require any permission (zero-permission) is installed on the victim device or a similar device.  
	\item The application is trained: some templates are created during the monitoring of a specific resource.  
	\item An algorithm is used to gather and analyse the obtained data.  
\end{itemize}

Table \ref{table1} summarises some of the most recent research on mobile side-channel attacks, discovered in the literature. The rows specify the inferred information such as visited websites, screen input or user locations whilst the columns define the side-channel leak source. \cite{b6} describe a proof-of-concept software, having limited permissions, which infers user-visited websites through analysis of incoming/outgoing network traffic. The network-traffic information is available through the \verb|/proc| filesystem (\verb|/proc/uid_stat|) as well as using the Android APIs. 
The authors performed the test cases using both sources for different test devices. Similarly, \cite{b12} monitored changes in the \verb|/proc/interrupts| system file following an approach called interrupt timing analysis and presented two new attacks to infer user unlock patterns and foreground apps on android devices. \cite{b11} created a proof-of-concept application which monitors hardware interrupts to infer approximately words entered by a user on the screen. Like the previous ones, it is based on values of \verb|/proc/interrupts| and \verb|/proc/stat|. A machine-learning classifier is employed to train and create a word fingerprint.  On the other hand, using a signature database, \cite{b16} proposed a method to infer the user running activities and activities transactions by collecting changing information from CPU, network and shared memory stored in the \verb|/proc| filesystem files (\verb|/proc/pid/stat|, \verb|/proc/uid_stat/uid/tcp_rcv|). This information is stored in a signature database and subsequently analysed to obtain the highest similarity result. The authors state that different Android versions have different impacts on side-channel issues. In fact, \cite{b7} propose a method to infer visited websites on Android devices by analysing power consumption data, which reflects the complexity and loading process of each webpage. Before Android 7.0, this information could be obtained accessing system files, requiring no specific permission (e.g. \verb|voltage_now| and \verb|current_now| in \verb|sys/class/power_supply/battery/|). However, starting from Android 7.0 this information cannot be collected anymore from non-root applications; therefore the power estimation method proposed in the paper is based on getting \verb|/proc| filesystem data about CPU load and CPU frequency. They are considered a good approximation of total power consumption, as they are the main influencing factors (\verb|/proc/stat|). However, starting from Android 8.0, most of the \verb|/proc| files are not accessible from android applications anymore. \cite{b4} presents one of the first project exploring operating system side-channel attacks on iOS devices. On this kind of device, it was challenging to get appropriate in-process statistics; instead, it was necessary to get global statistics and combine them using a specific ML algorithm to reduce the noise. The result of their work demonstrated that mobile operating system level side channel attacks are feasible on devices which do not have a procfs filesystem, such as iOS. The researchers were able to infer app in foreground, launched websites and map searches and this is also one of the first experiment using a timing-channel attack to infer information on mobile devices, since exploiting the {\sc fileExists} iOS API it was possible to verify the existence of a file, by monitoring the duration of the call to the function. Since most of the android sensors (such as accelerometer, magnetometer and gyroscope) do not require any permission to be queried for data, many researchers exploited these components to perform side-channel attacks. \cite{b14} presented a project to infer user input based on accelerometer and magnetometer data. Their work assumes that different user inputs generate different positions of the phone. They collected motion sensors holding the smartphone in three distinct positions and then used four types of algorithms to analyse the data and infer keyboard inputs. Similarly, \cite{b9} explains how to infer user ethnicity by collecting accelerometer reading on smartphones when a user is typing on a keyboard. They used a machine-learning algorithm called random forest for data analysis, which allowed to obtain accurate results and successfully classify which of the users were Chinese people. Furthermore, \cite{b10} proposed a method to infer user location by collecting sensors data on Android with an application having zero permission. Their approach consisted in transforming public map information into specific database structures; then sensor information was collected and sent to a server which analysed and correlated the structures (curvatures, accelerations and timestamps) to identify a matching route. Instead of using an application, \cite{b8} developed a project to track website visits of users using sensors. They used JavaScript to access accelerometer and gyroscope data, in order to fingerprint individual users, by employing a machine-learning technique to combine several features of the cited sensors and produce higher classification accuracy. A similar case of JavaScript usage to access sensors data was provided by \cite{b15}: in their work, they created {\sc TouchSignatures} which aimed to listen to orientation and motion sensors in order to identify user taps on the screen through a browser, by training a machine learning system with multiple classifiers. Finally, \cite{b13} recently developed an application called {\sc AlphaLogger}, which can infer pressed keyboard keys using motion sensors. The author states that their application provides better results than existing ones, and the combined use of magnetometer and accelerometer allows to obtain an accuracy of 90.2\%. The previously described projects are all based on a limited number of side-channel leak source (such as either specific sensors or specific procfs files), and no project covers all the types of events presented in Table \ref{table1}.  

\subsection{Automatic Side-channel analysis solutions.}
As described in the previous section, the discovery of side-channel vulnerabilities is usually a tedious and confusing task, and a form of automation is helpful to speed up the process and cover more sources or types of inferred information. However, to the best of the author’s knowledge, only two automatic solutions have been developed for the analysis of side-channel attacks on mobile devices. \cite{b3} presented {\sc ProcHarvester}, the first automatic scanning approach to discover procfs vulnerabilities on Android.  It monitors changes in multiple procfs files during the triggering of certain events, to infer their properties. However, as stated in the previous section, access to most of the procfs individual processes resources has been denied starting from Android 8. Its applicability was demonstrated by identifying new and already manually found information leaks. The approach is based on the concept of template attacks, and the benefits are that this methodology can be employed without having background knowledge of effects. They used the DTW algorithm to identify events correlations, which is often used to perform the comparison of misaligned time series, namely series which vary in time and speed \cite{b18}. The idea was to focus on events for which a correlation with time series could be observed to identify side-channel flaws successfully; a procfs resource was profiled if it changed with high frequency after the triggering of an event. The solution, however, only focuses on procfs system, leaving aside other types of side-channel sources (e.g. sensors). Furthermore, {\sc ProcHarvester} focuses only on three types of events: app starts, website launches and tap actions \cite{b3}. Similarly, \cite{b2} created a project called {\sc SCAnDroid}, to automate side-channel analysis of Android Java APIs. Most of the techniques follow an approach inspired by {\sc ProcHarvester}, and similarly, it employs the DTW algorithm to find correlations. Furthermore, it gathers a list of APIs from the official Android documentation or using Java reflection and tries to automatically select parameters for method calls. In this case, the solution monitored changes in return values after the execution of certain events. It was possible to generate APIs calls using Java reflection because the project was based on Android Java APIs and similarly to {\sc ProcHarvester} only supported the following events: website launches, google maps queries and applications start. {\sc SCAnDroid} covers most of the Android Java APIs but is limited to ones prefixed with the words get, is, has and query; all other relevant APIs (e.g. callback listener for sensors) are not covered by the methodology, because they are more complex to be invoked correctly and automatically. This also includes Android NDK APIs, which are not callable using Java reflection \cite{b19}. Moreover, {\sc SCAnDroid} was tested on Android 8.0, but at the time of writing, two other Android versions were released, adding additional methods to the codebase. Finally, {\sc SCAnDroid} profiling is based only on methods which return value changes subsequently to the triggering of an event, and the software could be extended to include additional types of attacks based on other factors, such as timing differences and non-original return values (e.g. the timing attack described by \cite{b4}). 

\subsection{Android SDK vs Android NDK} 
The Android SDK is a software development kit which allows developers to create Android application using the Android platform. It includes source codes, emulators and required libraries to build and run Android applications \cite{b20}. The Android applications are natively written using Java because android devices employ a custom virtual machine to run them called Dalvik Virtual Machine, which runs above the Linux kernel. A virtual machine is a software environment which implements abstract resources above the actual hardware being present. The Dalvik VM was explicitly designed for Android as in the past mobile systems had little RAM and low-performance CPUs. Also, the extended range of different architectures where Android can run made it necessary to add a layer of virtualisation to make it possible running generic applications on multiple devices. Therefore, it was necessary to have a virtual machine which running on limited resources could provide at the same time high performances \cite{b21}. However, a description of the Dalvik VM architecture is beyond the scope of this work. Due to the creation of a virtual machine to be the only way to run Android applications, the developers felt the need, in certain circumstances, to access physical devices components and other hardware features, which were not usable from Java code. That is why the Android NDK was introduced, a native development kit to use C and C++ code inside Android applications \cite{b22}. This is particularly useful when it is necessary to squeeze extra performance of the device, only accessible from native code, such as some specific hardware features or low-level libraries. Also, this made it possible to reuse already existing C and C++ libraries in Android projects, which was not possible to include in Java-based projects. The SDK and the NDK bring with them a set of Android platform libraries to access various capabilities of the Android operating system.

\subsection{Template Attacks}
A template attack is a specific side-channel technique to eavesdrop information from a device or software component, usually consisting of two phases: a profiling phase and attack phase. In order to perform a template attack, it is necessary to have access to a pair of similar devices (having, for example, the same model and capabilities). In the profiling phase, a training device is used to compute parameters, also called templates, for a component under test. Subsequently, in the attack phase, another device, called target, is tested using the generated templates, to infer some secret data from that component \cite{b23}. Some past projects demonstrated that using the same device for the profiling and attack phases could create incongruent results; however, this approach is followed by {\sc SCAnDroid}, {\sc ProcHarvester} and other projects \cite{b24}. The solution described in this work is based on the same approach, as a pair of similar devices was unavailable to the researcher. In {\sc ProcHarvester} and {\sc SCAnDroid}, template attacks are used to model (based on procfs reads or return values) and infer triggered events. An event can be whatever action is typically performed by a user on a smartphone, such as the launch of an application or a search on the browser.  

\subsection{DTW algorithm and k-fold cross-validation}
The {\em Dynamic Time Warping} (DTW) is an algorithm to measure the similarity between two temporal sequences which varies in speed. It is often used to recognise temporal sequences of video and graphics data as well as to perform speech and signature recognition. However, any data which can be turned into a linear sequence can be analysed using this algorithm. DTW allows comparing misaligned time series (for example, having different lengths) by finding the warping path with minimal distance between them \cite{b25}. In template attacks based on APIs, it is necessary to get the return value of some functions during a specific amount of time. Therefore a set of time-series is gathered for any triggered event, which consists of a combination of numerical values and timestamps; the time series form a so-called trace for a combination of method and event. To correctly identify the presence of a side-channel leak, the DTW algorithm is one possible solution to determine similarities between two or more traces. Other algorithms might exist to obtain better infer results, however, both {\sc SCAnDroid} and {\sc ProcHarvester} mentioned that for template attacks, among the algorithms tested, the one yielding the best results was the DTW. The DTW algorithm is used together with a technique called k-fold cross-validation; this is a resampling procedure approach which divides a set of data into k groups of equal size. This technique is used in applied machine-learning and focuses on a limited sample of data in order to estimate how a model generally performs, to make predictions on data which is not used in training \cite{b26}.  

\begin{figure*}
	\centering
	\includegraphics[width=1.0\linewidth]{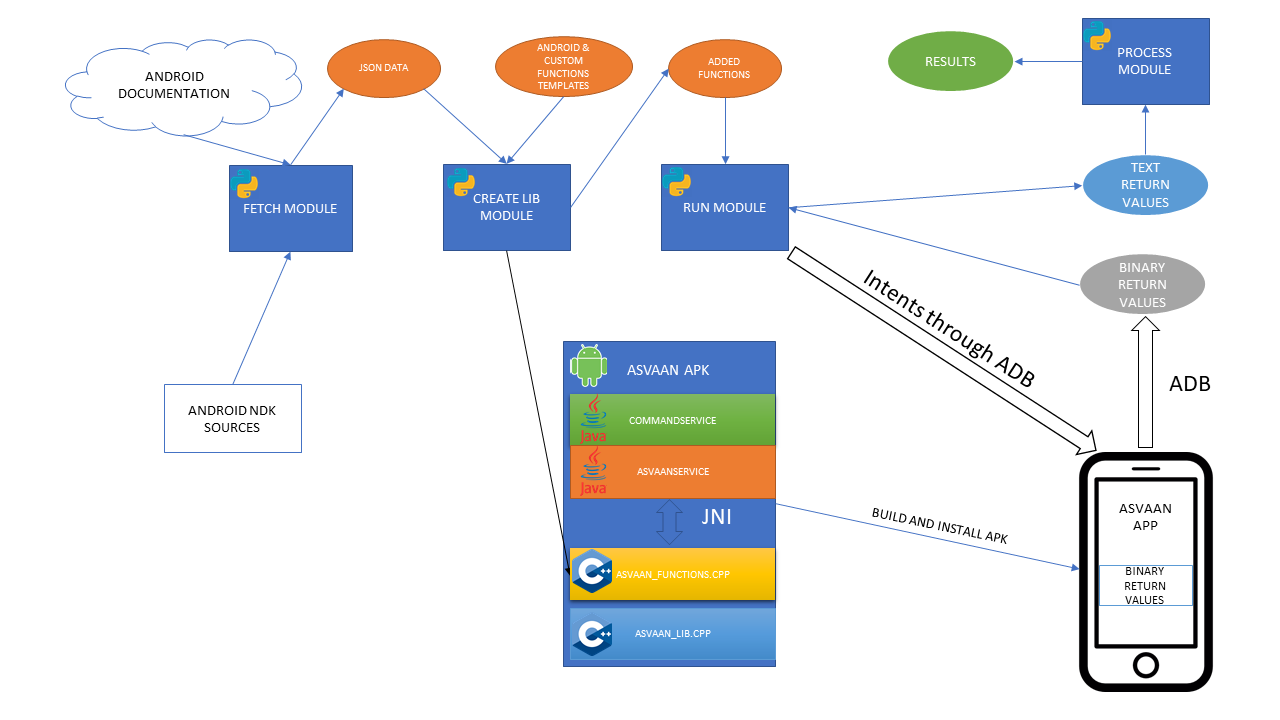}
	\caption{ASVAAN Architecture}
	\label{fig1}
\end{figure*}

\subsection{The problem of reflection}
{\em Reflection} is a property of software which allows to modify or examine the runtime status of programs and execute operations which have as object the program itself; this property allows to dynamically obtain, among the other things, the available methods, fields and classes at runtime. In other words, reflection is the ability to query runtime metadata of a program while it is running  \cite{b27}. Reflection is not a characteristic of every programming language, but usually only available in interpreted ones (such as Java) and not in compiled ones (e.g. C/C++). As defined by \cite{b28}, the most typical use of reflection is to create a new object, method invocation and perform fields access. \cite{b2} make use of reflection in {\sc SCAnDroid} to examine and correctly invoke the Java Android SDK methods at runtime, by retrieving their requirements (such as the necessary classes and methods). In their approach, they tested methods prefixed with specific keywords by inspecting packages and classes in the Android SDK, and for each one, they retrieved constructors and methods syntax, to invoke them with valid parameters. This technique permitted to scan the entire Android SDK bundle without prior knowledge of methods semantic. On the other hand, they also reported that some called methods crashed during the tests on the Android 8.0 codebase, due to invalid parameters passed as arguments for methods. Since the research is based on testing the Android NDK, only native methods are investigated, and no reflection is available, as these methods are implemented using C/C++. Therefore, it was necessary to parse the available methods from another source, such as the Android NDK developer page and the NDK source code, and subsequently, think about an approach to call them at runtime dynamically. Furthermore, reflection could also have made it easier to create valid objects at runtime as arguments for method calls. The problem of creating a valid call to test Android native libraries is currently a hot research topic which was very recently explored by \cite{b29} for fuzzing purposes. Fuzzing is a popular technique to discover vulnerabilities in software which in most cases consists in performing a call to functions with invalid, unexpected and random arguments. In their paper, they describe a way to infer possible values and corresponding types for function arguments using data flow analysis, by assigning different attributes to each argument based on some observations. However, due to the complexity of this approach, this is beyond the scope of this research. Although the research initially was thought to use a fully-automatic scanning technique, the reflection problem made the creation of valid arguments for functions an infeasible task at the moment. Since reflection cannot be used in the current context,  the researcher opted for a semi-automatic solution where a hypothetic user has to define a valid call for a function in a specific template file saved in a folder. When the templates are correctly implemented for the functions under test, the solution performs an automatic analysis as initially described. 

\section{ASVAAN}
{\sc ASVAAN} is the name given to the solution described in this work to semi-automatically scan the Android NDK (APIs and other native libraries functions). It consists of two macro sections: four python components and an android application (the latter having both native and Java code). These two sections communicate using the ADB bridge and Android Intents, to send commands and gather return values of functions. {\sc SCAnDroid} highly inspired many parts of the {\sc ASVAAN}'s architecture, but some component had to be reinvented for the considerations raised in the previous section. 
The solution architecture is shown in Figure \ref{fig1}. First, a user runs the python fetch component to gather NDK methods from the Android NDK developer documentation and cross-correlate them with the Android NDK sources. The methods are parsed by the fetch component, according to a set of rules defined in a configuration file, and a JSON data file is created. Then, the user launches the create lib python component, which uses the data file generated by the fetch module and user-defined function templates, to dynamically generate a native NDK library containing a proper call for the functions under test. After generating the native library, the Android application is compiled and installed on the physical test device (connected to the computer through a USB cable). When the application is installed, the user uses the run python component to trigger a new testing phase, which consists in simultaneously logging function return values and launching events of a specific type (such as application launches, website launches and maps queries). In this phase, the python run component instructs the android application by generating some intents and sending them with ADB, which are then received and parsed from an Android Intent service. Services are components of Android applications useful to perform long-running operations in the background of the Android OS, without providing a user interface. Once started, services can run either in the background or foreground until the user stops them. An intent service is a useful subclass of Android services which handles requests expressed as Intents. When the service is started, it waits for new Intents and parses its parameters \cite{b30}. After receiving an intent, the application communicates with its native libraries using JNI, which performs the actual logging of the traces for the tested functions, subsequently saving them on specific binary files. JNI is an interface which permits the Java layer to communicate with native code and libraries bidirectionally. This process is accomplished by passing parameters to native functions in order to obtain, after the invocation, a result back from Java \cite{b31}. At the end of this phase, the binary files are pulled from the android filesystem device into a folder and transformed into text files by the run component. In the final step, the process library component is used to apply the k-fold cross-validation and the DTW algorithm on the obtained traces, in order to verify the presence of side-channel leaks. The implementation of this component is directly taken from {\sc SCAnDroid}. 

Each component of the solution is described more in detail in the subsequent sections of this chapter. 

\subsection{Fetch Component}
The fetch component is a python script which parses data from the Android NDK developer documentation website and correlates them with the Android NDK sources, which must be locally available on the testing computer. The data obtained is parsed and saved on a data file for later usage by the create lib component. The correlation is necessary to remove methods which are parsed from the documentation but might not be available for a specific API level. This information is incomplete on the NDK developer documentation website; therefore, it is easier to parse it through the source files of the Android NDK bundle, as functions which are introduced only from a specific API level are marked with the following macro: 

\texttt{\_\_INTRODUCED\_IN(<API\_LEVEL>)}

The parsing process is based on the following steps: 
\begin{itemize}
	\item The NDK developer webpage is downloaded, and the script parses all the functions and imports related to the available modules.   
	\item The parsed data is correlated with the NDK sources. This part of the script is configurable by editing a configuration file containing a set of rules.    
	\item The component creates a new JSON file (which is then parsed by the creating lib component) containing the functions which need to be tested and included in the dynamic android application library.  
\end{itemize}

The initial idea was to parse all the functions (also the ones not specified in the developer documentation and belonging to native libraries) directly from the Android NDK sources. However, the idea was abandoned, because it was encountered the problem of successfully parsing C++ code structures using regex, caused by the complexity of the programming language. 

\subsection{CreateLib Component}
The create lib component uses and parses the file created by the fetch component to create a dynamic Android NDK library containing the calls for the selected functions. Data is parsed in a specific order. First, the imports necessary to call the functions are parsed from the JSON file and written at the start of the dynamic library. Also, some default imports are included. Then, for each function parsed from the data generated by the fetch component, a function stub is incrementally numerated and created. These stubs are called by a specific function at runtime, depending on the value of a passed argument. The stub is written to the native file, inside a new function, created according to the format described in Listing \ref{lis1}. The functions implementations are taken from templates text files declared in a folder. As specified in the research background chapter, this is the only section which needs to be compiled explicitly by the software user, because of the described reflection problem. Also, templates files are divided into Android and custom. Android functions templates are necessary to successfully implement the functions gathered by the fetch component and consist of performing a clean call of that function and returning the return value specified in the android developer documentation.

Custom functions templates are created by the user, using any combination of function, argument and return value, including functions of other native libraries.  

\begin{lstlisting}[caption={Functions stub},captionpos=b,frame=single,label={lis1}]
 <RETURN VALUE> Function_<FUNCTION NUMBER>(){
	<STUB>
}
\end{lstlisting}

This possibility was made available in order to provide an answer to the third question posed by the research, which is based on combining return values with other function behaviours to identify additional sources of leaks. Custom functions templates files have the format described in Listing \ref{lis2}

Finally, the component generates the call\_function function which calls a single or multiple functions stubs, depending on the command passed by the run component. The output of the component consists of the creation of a native NDK library which is built and bundled in the APK file; it is then used by the {\sc ASVAAN} Android services. After having created the library, a file is created containing the total number of written functions, in order to communicate this information to the next component.  

\begin{lstlisting}[caption={Custom functions templates},captionpos=b,frame=single,label={lis2}] 
RV=<RETURN VALUE>
DEP=<DEPENDENCY HEADERS>
STUB
<FUNCTION STUB RETURNING> 
ENDSTUB
\end{lstlisting}

\subsection{Run Component}  
The run component is the main python component and the only point of contact with the Android application; among the other things, it allows to start and stop the logging of tested functions. To perform this action, it uses ADB to generate new intents and communicate with the Android application installed on the device.
 
The run component works in three different events: application launches, maps queries and websites launches. Two of the three the events are described more in detail in the testing section, but the component can be easily extended with other ones.
 
After having selected the scenario, the component parses the total number of functions defined in the text file created by the create lib component and starts the logging phase. The user can select which functions to test among the one created (for example, all, only one or multiple functions) by passing a list of values as an argument to the component. In the logging phase, for each function, the component randomly selects an event value from the ones defined in the configuration file and triggers the functions recording by communicating with the android application. The logging process for each event goes on for a user-configurable number of seconds, and then it is stopped. After the process is finished, the files containing the functions traces are pulled from the android filesystem and converted to text-format, for later analysis by the process component. The traces binary files have the format illustrated in Listing \ref{lis3}.

\begin{lstlisting}[caption={Traces binary files format},captionpos=b,frame=single,label={lis3}]
	<RETURN VALUE SIZE>
	<RETURN VALUE 1>
	<TIMESTAMP 1>
	<RETURN VALUE 2>
	<TIMESTAMP 2>
	...
	...
	...
	<RETURN VALUE N>
	<TIMESTAMP N>
\end{lstlisting}

\subsection{Process Component}  
The process component parses the files containing the traces to infer the probability of an information leak for a combination of event and function. It reuses the algorithm implemented in {\sc SCAnDroid}, consisting of a combination of DTW and k-fold cross-validation. Most of the component implementation is similar to the  {\sc SCAnDroid} processing section. The reuse is allowed respecting the license of the solution, as the Zlib license is compatible with the GPLv3 \cite{b32}. The component performs the following steps: 
\begin{itemize}
	\item Parses function traces from the text files and adapt them to a usable format.    
	\item Performs k-fold cross-validation on the traces.    
	\item Takes each fold and computes the warping path with minimal distance between the values in the fold and all other traces. 
\end{itemize}

\subsection{Android Services} 
The Java section of the Android application is necessary to receive and parse intents from the run python component; when an intent is received, it instructs the native part to call a specific function or multiple functions. The Java section is mostly composed of Android services. These services run in the foreground and are mainly responsible for calling the native substrate to log functions return values on the Android filesystem during the logging phase. The services receive start and stop events commands from the run component to control the logging phase. The functions to test in a single logging iteration are specified as argument of an Intent sent by the run component. In order to call the native libraries, the Java section defines some JNI functions, which modify and retrieve the state of native structures and variables. 

\subsection{Android Native Libraries}
The native libraries section is responsible for writing the return value of functions to specific files in the android filesystem, for later retrieval and analysis. The process is performed in the following way, depending on the data received by the Android services:  
\begin{itemize}
	\item Perform the call of the function/functions in exam    
	\item Continuously write the return value and timestamp on a specific binary file in a publicly accessible folder of the android filesystem (e.g. \verb|/sdcard|). 
	\item Handling some types of native crashes and in case, inform the run component to stop the logging or pass to the next function.   
\end{itemize}

\begin{figure*}
	\centering
	\includegraphics[width=1.0\linewidth]{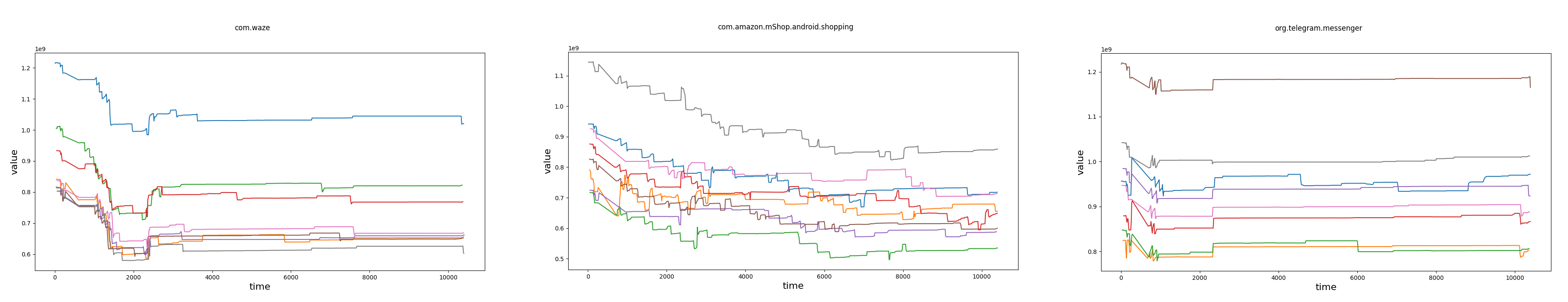}
	\caption{Traces gathered from function sysinfo(1) launching the Waze, Amazon and Telegram application, respectively.}
	\label{fig2}
\end{figure*}

\section{Evaluation}
In this section, it is demonstrated the real-world applicability of {\sc ASVAAN} in two different scenarios: application launches and website launches. A detailed description of each of these scenarios is illustrated in the following sections. The prototype was evaluated by discussing the new side-channel information leaks found in the Android 10.0 NDK code base; the vulnerabilities found were responsibly reported to Google as part of their ASR program. Physical devices were preferred over emulator due to the need for accuracy in gathering traces from tested functions, which could be faked on emulators. Table \ref{table2} reports the smartphone Android devices used for testing. 

\begin{table}[htb]
	\caption{Test devices}
	\begin{center}
		\begin{tabular}{|l|l|l|l|}
			\hline
			{\bf Device}  & {\bf Model} & {\bf Android version} & {\bf API level} \\ \hline
			Galaxy S20+   & Samsung & Android 10.0  & 29    \\ \hline
			Galaxy S8+ & Samsung  & Android 10.0  & 29  \\ \hline
		\end{tabular}
		\label{table2}
	\end{center}
\end{table}

As mentioned in the previous sections, initially the solution was intended to be fully automatic. However, during the research, due to the reflection problem, it was necessary to make the user performing some manual actions to create the NDK functions templates. Due to time constraints, it was not possible to create valid function templates for all the functions in the NDK codebase, so only a part of them was tested. However, the tested functions were enough to find new information leaks and provide valid evaluation results for the followed methodology. Following are reported the results of the testing for each of the scenarios considered. Since the results are similar between the two devices, only the ones belonging to the Galaxy S20+ are reported in this evaluation.

\begin{table}[htb]
	\centering
	\caption{Application launches events}
		\begin{tabular}{|l|}
			\hline
			{\bf Application launches event}   \\ \hline
			com.waze  \\ \hline
			com.whatsapp  \\ \hline
			org.mozilla.firefox   \\ \hline
			com.instagram.android  \\ \hline
			com.paypal.android.p2pmobile   \\ \hline
			com.tripadvisor.tripadvisor   \\ \hline
			com.facebook.katana   \\ \hline
			com.netflix.mediaclient   \\ \hline
			com.amazon.mshop.android.shopping   \\ \hline
			org.telegram.messenger   \\ \hline
			com.linkedin.android   \\ \hline
			com.twitter.android   \\ \hline
			com.ebay.mobile    \\ \hline
			com.booking \\ \hline
			com.snapchat.android \\ \hline
		\end{tabular}
		\label{table3}
\end{table}

\subsection{Application Launches} 
This event consists of starting a new application installed on the device from the home screen. Information about started applications should be kept secret, as starting from Android 5.0 is not possible for packages to get a list of applications currently running on an Android device \cite{b2}. The application launches were randomly performed using the \verb|adb monkey| command.  Table \ref{table3} reports some popular Android applications downloadable from the Play Store, which were tested for this scenario. 

The functions were recorded for ten seconds for each application launch, gathering eight series of traces per event.

Table \ref{table4} depicts the top four functions having higher information leak accuracy for the tested application launches taken from the pool listed in Table \ref{table3}; The highest probability inference is given from custom functions call defined by the researcher (which stub call implementation is reported in Appendix \ref{appendix:a}).  Figure \ref{fig2} shows the eight traces gathered for the highest inference accuracy function, related to the following three applications: {\sc Waze}, {\sc Amazon} and {\sc Telegram}. It is possible to notice that traces in the same graphics have similar characteristics and are distinguishable from traces in other graphics representing different events. The inference accuracy represents how accurate was {\sc ASVAAN} in identifying similar traces using the DTW and the k-fold cross-validation.  

 \begin{table}[htb]
	\caption{Application launches infer accuracy results}
	\begin{center}
		\begin{tabular}{| l | c | c | c |}
			\hline
			{\bf Function} & 	{\bf Location} & {\bf Accuracy} & {\bf Status}   \\ \hline
			sysinfo(1) & \verb|sys/sysinfo.h| & 73\% & Won't Fix   \\ \hline
			statvfs & \verb|sys/statvfs.h| &  73\% & Won't Fix   \\ \hline
			get\_avphys\_pages& \verb|sys/sysinfo.h| & 61\%& Won't Fix   \\ \hline
			sysconf & \verb|unistd.h| & 60\% & Won't Fix   \\ \hline
		\end{tabular}
		\label{table4}
	\end{center}
\end{table}

The results of the test allowed to discover new side-channel leaks in functions of the Android NDK codebase.

\begin{figure*}
	\centering
	\includegraphics[width=1.0\linewidth]{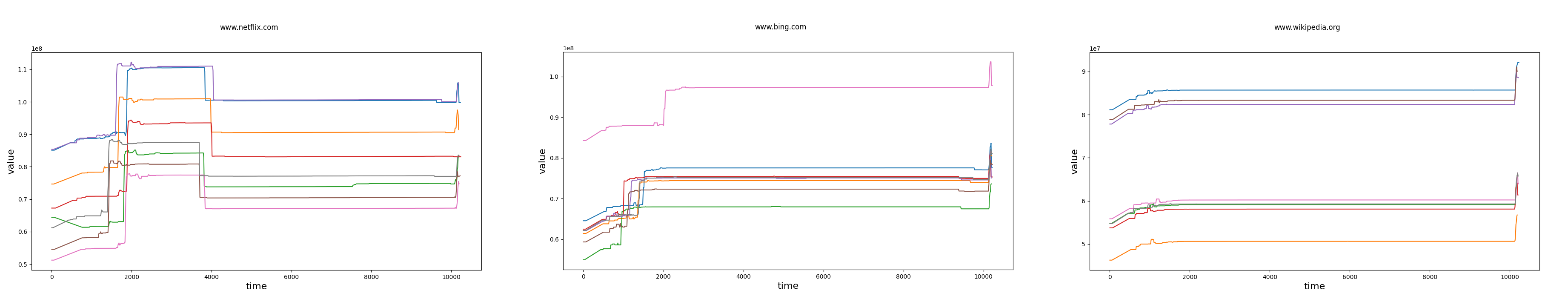}
	\caption{Traces gathered from function sysinfo(2) launching Netflix.com, Bing.com and Wikipedia.org, respectively.}
	\label{fig3}
\end{figure*}

\subsection{Website Launches} 
A website launch consists of browsing on a specific webpage using the Google Chrome browser. Querying the user history from an application package was disabled since Android M \cite{b2}. The website launches were randomly performed using the \verb|adb am start command|. Table \ref{table5} reports some popular websites (according to alexa.com) which were tested for this scenario. 

\begin{table}[htb]
	\caption{Website launches events}
	\begin{center}
		\begin{tabular}{|l|}
			\hline
			{\bf Website launches event}   \\ \hline
			http://www.facebook.com   \\ \hline
			http://www.google.it  \\ \hline
			http://www.yahoo.com    \\ \hline
			http://www.bing.com  \\ \hline
			http://www.aruba.it   \\ \hline
			http://www.imgur.com   \\ \hline
			http://www.baidu.com   \\ \hline
			http://www.wikipedia.org   \\ \hline
			http://www.sohu.com   \\ \hline
			http://www.gmail.com  \\ \hline
			http://www.tmall.com   \\ \hline
			https://www.amazon.com   \\ \hline
			https://www.taobao.com    \\ \hline
			https://www.netflix.com \\ \hline
			https://twitch.tv \\ \hline
			https://qq.com \\ \hline
		\end{tabular}
		\label{table5}
	\end{center}
\end{table}

As with the previous scenario, the functions were recorded for ten seconds for each website launch, gathering eight series of traces per event. 

\begin{table}[htb]
	\caption{Websites launches infer accuracy results}
	\begin{center}
		\begin{tabular}{| l | c | c | c |}
			\hline
			{\bf Function} & 	{\bf Location} & {\bf Accuracy} & {\bf Status}   \\ \hline
			sysinfo(2)  & \verb|sys/sysinfo.h| & 85\% & Won't Fix  \\ \hline
			fopen (1) &\verb|stdio.h| & 71\%  & Won't Fix \\ \hline
			fopen (2) & \verb|stdio.h| & 62\% & Won't Fix \\ \hline
			fopen (3) & \verb|stdio.h| &52\% &  Won't Fix \\ \hline
		\end{tabular}
		\label{table6}
	\end{center}
\end{table}

Table \ref{table6} depicts the top four functions having higher information leak accuracy for the tested website launches taken from the pool listed in Table \ref{table5}. As it is possible to notice from the table, similarly to the previous case, the highest probability inference is given from custom functions calls.  

Figure  \ref{fig3} shows the eight traces gathered for the highest inference accuracy function, launching the following three websites: {\sc Netflix.com}, {\sc Bing.com} and {\sc Wikipedia.org}; similarly to application launches, the traces in the same figure are clearly different from traces in other figures. Furthermore, as in the previous case, the resulting information leaks were responsible reported to the Android Security Team.

\section{Conclusion}
This research introduced a semi-automatic solution to scan Android NDK APIs and functions of native libraries for side-channel vulnerabilities, which could allow inferring some events of interests such as application launches and website launches. A methodology to scan Android NDK APIs is necessary, as according to the literature review, existing similar research only focus on procfs information leaks and Android Java APIs. Android NDK and native libraries, which are a large part of the Android codebase, were left entirely aside from existing solutions; also, projects directly targeting the NDK for side-channel leaks have not been found in the available literature. Furthermore, a similar solution is necessary to successfully identify and fix information leaks which could be abused by malicious actors to steal sensitive information from Android devices. The evaluation was applied to the Android 10.0 codebase and successfully confirmed new information leaks in functions belonging to Android native libraries. Most of the vulnerable functions allow to query variable system information such as available RAM or number of processes currently running on the device, and most of them are part of the standard C library of the Android operating system (Bionic). Furthermore, not all the events could be inferred from the vulnerable functions: some of the functions allowed to infer application launches and website launches with acceptable accuracy, however, no information leak was found related to inference of map queries. This does not confirm that in the NDK codebase does not exist a method which could allow inferring that type of scenario, but successfully inferring two types of events out of three was considered an excellent compromise to demonstrate the applicability and importance of the methodology described in the research.

\subsection{Limitations and future work }
Although the initial idea was to create a fully automatic solution, the need to work with C functions, which arguments semantic and arguments dependencies was unknown at runtime, made it necessary to introduce user-created data to run the solution properly. In order to solve this problem, the concept of function templates was introduced, which once defined by the users with valid calls, allowed to scan the tested functions for side-channel vulnerabilities automatically. As a consequence, one of the most critical limitations of the created software is partial code-coverage of Android NDK: it was possible to test only a part of the available functions, as only a portion of the functions templates were created due to time constraints. The tested functions were enough to find new information leaks and perform an evaluation, but this does not guarantee that other information leaks could be present in not tested functions.  

Furthermore, initially, the methodology was thought to be only adapted to Android NDK APIs. However, to expand its potential, it was introduced the concept of custom function templates, which allowed to test functions belonging to NDK native libraries not listed on the Android NDK developer documentation. The initial idea was to parse the syntax of those functions from the source. However, due to the complexity of parsing C/C++ sources, at the moment, it is necessary to create those templates by manually finding the functions.  

As mentioned in the previous sections, the problem of automatically creating a valid call to test android native libraries functions was investigated by \cite{b29}. However, due to its complexity, it was left aside from this research; to further improve the automatism of the methodology, their findings could be adapted to the solution in order to automatically create valid calls to Android NDK functions.  

Finally, most of the existing works about side-channel vulnerabilities on mobile devices is based on the Android operating system. At the best of author knowledge, few works  \cite{b4} and more importantly, no automatic side-channel analysis solutions were created for other mobile operating systems. For example, iOS, similarly to Android, is based on Linux and has a large codebase which could be investigated for information leaks. Future works could be focused on different mobile operating systems to close the gap in the lack of automated software solutions to find information leaks on mobile devices.

\clearpage

\appendix

\section{Functions templates}
\label{appendix:a}
Following, are reported the stub templates used for the NDK functions tested as part of the evaluation performed in the relative section.

\begin{lstlisting}[caption={sysinfo(1)},captionpos=b,language=C,breaklines,frame=single]
	struct sysinfo si; 
	sysinfo(&si);
	return si.procs; 
\end{lstlisting}

\begin{lstlisting}[caption={statvfs},captionpos=b,language=C,breaklines,frame=single]
	struct statvfs st;
	int i = statvfs("/sdcard",&st); 
	return st.f_bavail;  
\end{lstlisting}

\begin{lstlisting}[caption={get\_avphys\_pages},captionpos=b,language=C,breaklines,frame=single]
	return get_avphys_pages(); 
\end{lstlisting}

\begin{lstlisting}[caption={sysconf},captionpos=b,language=C,breaklines,frame=single]
	return sysconf(_SC_AVPHYS_PAGES); 
\end{lstlisting}

\begin{lstlisting}[caption={sysinfo(2) },captionpos=b,language=C,breaklines,frame=single]
	struct sysinfo si; 
	sysinfo(&si);
	return si.sharedram;
\end{lstlisting}

\begin{lstlisting}[caption={fopen(1)},captionpos=b,language=C,breaklines,frame=single]
	FILE *file; 
	long pages = 0; 
	char test[20];
	if(file = fopen("/sys/class/net/wlan0/statistics/tx_bytes", "r")){ 
		fgets(test, 20, file);
		fclose(file);
	}
	pages = atol(test); 
	return pages;
\end{lstlisting}

\begin{lstlisting}[caption={fopen(2)},captionpos=b,language=C,breaklines,frame=single]
	FILE *file;
	long pages = 0; 
	char test[20];
	if(file = fopen("/sys/class/net/wlan0/statistics/rx_packets", "r")){
		fgets(test, 20, file);
		fclose(file);
	}
	pages = atol(test); 
	return pages;
\end{lstlisting}

\begin{lstlisting}[caption={fopen(3)},captionpos=b,language=C,breaklines,frame=single]
	FILE *file;
	long pages = 0;
	char test[20];
	if(file = fopen("/sys/class/net/wlan0/statistics/tx_packets", "r")){
		fgets(test, 20, file); 
		fclose(file);
	}
	pages = atol(test);
	return pages;
\end{lstlisting}
	
\end{document}